\documentclass[preprint,12pt, a4paper]{elsarticle}

\usepackage{amsmath,amsbsy}
\usepackage{hyperref}

\usepackage{subfig}
\biboptions{round, authoryear}

\usepackage{float}
\usepackage{amsmath}

\journal{Astronomical Journal}

\begin{document}

	\begin{frontmatter}
	
		\title{A Dynamic Trajectory Fit to Multi-Sensor Fireball Observations}
        
		\author{Trent Jansen-Sturgeon\corref{cor1}}
 		\author{Eleanor K. Sansom\corref{cor1}}
  		\author{Hadrien A. R. Devillepoix\corref{cor1}}
  		\author{Philip A. Bland\corref{cor1}}
  		\author{Martin C. Towner\corref{cor1}}
  		\author{Robert M. Howie\corref{cor1}}
 		\author{Benjamin A. D. Hartig\corref{cor1}}

		\cortext[cor1]{School of Earth and Planetary Sciences, Curtin University, GPO Box U1987, Perth, WA 6845, Australia}
		
		
		\begin{abstract}
		    Meteorites with known orbital origins are key to our understanding of Solar System formation and the source of life on Earth. However, these pristine samples of space material are incredibly rare. Less than 40 of the 60,000 meteorites held in collections around the world have known dynamical origins. Fireball networks have been developed globally in a unified effort to increase this number by using multiple observatories to record, triangulate, and dynamically analyse ablating meteoroids as they enter our atmosphere. The accuracy of the chosen meteoroid triangulation method directly influences the accuracy of the determined orbit and the likelihood of possible meteorite recovery.

            There are three leading techniques for meteoroid triangulation discussed in the literature: the Method of Planes, the Straight Line Least Squares method, and the Multi-Parameter Fit method. Here we describe an alternative method to meteoroid triangulation, called the Dynamic Trajectory Fit. This approach uses the meteoroid's 3D dynamic equations of motion to fit a realistic trajectory directly to multi-sensor line-of-sight observations. This method has the ability to resolve fragmentation events, fit systematic observatory timing offsets, and determine mass estimates of the meteoroid along its observable trajectory.

            Through a comprehensive Monte-Carlo analysis of over 100,000 trajectory simulations, we find this new method to more accurately estimate meteoroid trajectories of slow entry events ($<$25\,km/s) and events observed from low convergence angles ($<$10$^{\circ}$) compared to existing meteoroid triangulation techniques. Additionally, we triangulate an observed fireball event with visible fragmentation using the various triangulation methods to show that the proposed Dynamic Trajectory Fit implementing fragmentation to best match the captured multi-sensor line-of-sight data.
		\end{abstract}
		
		
	\end{frontmatter}


\section{Introduction}\label{sec:intro2}
Fireball networks have been around since the 1960's with the specific goal of observing meteors from multiple stations to determine their past and future trajectories \citep{ceplecha_multiple_1961}. The meteoroids of real interest are the bright, deeply penetrating kind, with the highest chance of surviving the violent atmospheric entry process to produce meteorites. Finding meteorites with known orbits is key for giving these cosmic samples a regional context in the greater Solar System, potentially helping to answer some of the biggest questions in planetary science, such as the Solar System formation, and the origin of life on Earth. As of mid-2019, only 36 out of about 60,000 collected meteorites have known orbits; 5 of which have been found in recent history by the Global Fireball Observatory (GFO), a global collaboration of fireball networks, including Australia's Desert Fireball Network (DFN).

Successful recovery of the incoming meteorite requires accurate knowledge of the fall position. If found, it is highly desirable to have a well constrained and accurate orbit associated with the sample. Determining an orbit requires the entry radiant and velocity of the meteoroid, while prediction of fall positions require darkflight modelling, where darkflight is the period of meteoroid free fall to Earth after visible observations cease, during which the body is strongly influenced by its size and shape, as well as atmospheric winds. At the heart of all this dynamic analysis lies the triangulation and modelling of the observed luminous trajectory; giving both the darkflight and orbit determination their initial conditions. To improve the accuracy of these predictions, we must first improve the accuracy of our triangulation modelling techniques.

Three prominent methods of meteoroid triangulation have been documented and used in the past; Method of Planes \citep{ceplecha_geometric_1987}, Straight Line Least Squares \citep{borovicka_comparison_1990}, and Multi-Parameter Fit \citep{gural_new_2012}. These three methods are outlined conceptually below. For more detail and mathematical rigour, please refer to their respective papers.

A notable additional technique is the particle filter modelling method of \citet{sansom_3d_2019} as an alternative to the traditional triangulation methods. While particle-type approaches are thorough, they are also quite computationally intense and are not feasible as the default triangulation method for large meteoroid data-sets. Instead, it is generally best suited to special cases when a surviving meteorite is suspected.

\subsection[Method of Planes]{Method of Planes \citep{ceplecha_geometric_1987}}\label{ssec:mop}
Although the Method of Planes (MOP) is the oldest and least accurate of the three prominent triangulation methods, it is very computationally simple and often used for constructing the initial trajectory guess for more complex methods, such as the Straight Line Least Squares and Multi-Parameter Fit (see Section~\ref{ssec:slls} and Section~\ref{ssec:mpf}). MOP comprises four main steps: plane construction, radiant formation, position determination, and velocity fitting. 

To begin, MOP constructs a plane for every sensor. This plane includes the sensor's location and best fits to its associated observation rays using a least squares approach. It does so by adjusting the plane normal to minimise the square of the angular residuals between the rays and the plane.

Once the optimum plane is calculated for each sensor, they are intersected in 3D space to determine the straight line trajectory. In the case where more than two sensors recorded the meteoroid, a statistical weighting can be used to combine the straight line solutions from every different sensor-pair combination to produce one unique straight-line trajectory. This weighting is based on the convergence angle between the two planes as well as on the combined angular span of the observed meteoroid across the sensors.

Positions along the determined straight-line trajectory are found for every observation (regardless of time) as the closest point on the trajectory-line from that observed line-of-sight. These 3D positions are generally calculated by the intersection of the trajectory itself with a series of planes that each contain an individual line-of-sight and its associated optimum plane normal.

Lastly, the velocities are determined by fitting a model to the positional lengths along the trajectory as a function of time. These velocity models and fitting methods are described by \citet{pecina_new_1983, pecina_importance_1984}. However, it is interesting to note that \citet{pecina_new_1983, pecina_importance_1984} state these equations are ``violated" for longer trajectories, indicating the simplicity of their chosen velocity models.


\subsection[Straight Line Least Squares]{Straight Line Least Squares \citep{borovicka_comparison_1990}}\label{ssec:slls}
Only three years following MOP, the Straight Line Least Squares (SLLS) method was published. Although \citet{borovicka_comparison_1990} showed the SLLS method to produce lower residuals than MOP, they concluded that both methods produce similar results and could not recommend one over the other; even suggesting a combination of both may be preferable, depending on the case. That said, \citet{gural_new_2012} found the SLLS method to be more robust when lower resolution cameras were used. 

Unlike MOP, the SLLS method best fits a straight line trajectory directly to all the observed lines-of-sight at once. It does so by minimising the perpendicular distances between the lines-of-sight and the straight line trajectory itself. It was later stated by \citet{gural_new_2012} that a better alternative to the initially published SLLS method was to minimise the angular distance rather than the perpendicular distance. Using the angular distance acts to indirectly weight the line-of-sight measurements based on their observation range.

The positions are determined for every line-of-sight by determining the closest point on the optimised straight line trajectory to that given line-of-sight (regardless of time). Similar to MOP, the SLLS method requires a separate step to determine the velocity along the trajectory. The methods of \citet{pecina_new_1983, pecina_importance_1984} are used to determine this velocity by considering the 1D lengths along the trajectory over time.

We must note that \citet{borovicka_comparison_1990} offers the SLLS method in both the Earth Centered/Earth Fixed (ECEF) frame and the Earth Centred Inertial (ECI) frame; the main difference is where the straight line trajectory is defined. Performing the SLLS method in the ECI frame implicitly includes the Coriolis force, but requires absolute timing knowledge to operate. It is up to the user to determine which variation is more physically realistic.


\subsection[Multi-Parameter Fit]{Multi-Parameter Fit \citep{gural_new_2012}}\label{ssec:mpf}
The previously discussed MOP and SLLS methods are purely geometric triangulation solutions; i.e. the trajectory fitting component can be performed without any timing information. It is only as a second step, when velocity analysis is needed, that timing of the observed meteoroid is considered. This means MOP and SLLS determine a unique position for every line-of-sight; if there are simultaneous observations from N sensors, there will be N unique positions along the trajectory corresponding to the same point in time. Only later, in the velocity analysis step, can this potential scatter be dealt with.

The Multi-Parameter Fit (MPF) technique of \citet{gural_new_2012} differs from the previous two triangulation methods in that it fits raw observations directly to a trajectory solution, combining the straight line fitting and velocity modelling steps into one. Hence, N simultaneous observations will now result in one unique position along the trajectory. One implication of this approach is that the convergence angle can now be thought of as the angle between simultaneous lines-of-sight rather than between planes, which is a significant distinction.

As the name suggests, the MPF algorithm best fits unknown trajectory parameters to the measured lines-of-sight by minimising the angular distance between said lines-of-sight and the predicted lines-of-sight given their modelled positions along a straight line trajectory. These fitting parameters include the initial position ($\vec{p}_0$), the initial velocity ($\vec{v}_0$), some deceleration coefficients depending on the chosen model ($a_i$), and sensor timing offsets ($\Delta t_k$), giving the MPF the ability to handle asynchronous sensors assuming they all have relative timing. Positions along the straight line trajectory are determined using one of three velocity models: a constant velocity along the track, a linearly decreasing velocity with time, or an exponentially dependent deceleration \citep{jacchia_analysis_1967}. However, these suggested velocity models do not physically represent the trajectory dynamics. \citet{gural_new_2012} suggests that this technique is most applicable to smaller mass meteors ($<5$\,g) of short duration ($< 3$\,sec), unless a better model is used. 


\section{New Approach - Dynamic Trajectory Fit}\label{sec:dtf}
Of the three most prevalent triangulation methods (as discussed in Section~\ref{sec:intro2}), none claim to be able to fit long-duration fireballs of significant mass, in part because all methods have assumed a straight line trajectory. While \citet{jenniskens_meteor_2006} claim that masses $<50$\,g or equivalent magnitude of -2 can be approximated using straight line trajectories, if the goal is to observe deeply penetrating fireballs such as those targeted by the GFO, the fireballs are not guaranteed to follow this straight line assumption. In fact, \citet{sansom_3d_2019} show that the straight line assumption is an oversimplification that will affect orbit calculations and meteorite search regions for a significant number of fireball events.

The Dynamic Trajectory Fit (DTF) method proposed here removes this straight line assumption by fitting differential equations of motion directly to the measured lines-of-sight, thereby including all spacial/temporal information in one step and ultimately providing a more realistic account of the meteoroid's fall trajectory. This methodology takes the ideas of global fitting proposed in \citet{gural_new_2012} several steps further. The differential equations that describe meteoroid fall dynamics and ablation are as follows \citep{sansom_novel_2015}: 

\begin{equation}\label{eq:dv_orig}
    \frac{d\vec{v}}{dt} = - \frac{c_d A \rho_a \big\|\vec{v}_{rel}\big\|}{2 \rho_m^{2/3} m^{1/3}} \vec{v}_{rel} + \vec{a}_{grav}
\end{equation}
\begin{equation}\label{eq:dm_orig}
    \frac{dm}{dt} = - \frac{c_h A \rho_a m^{2/3} \big\|\vec{v}_{rel}\big\|^3}{2 H^* \rho_m^{2/3}}
\end{equation}

where $\vec{v}$ is the meteoroid's absolute velocity in the Earth-Centred Inertial (ECI) frame, $\vec{v}_{rel}$ is the meteoroid's velocity relative to the atmosphere, $m$ is the meteoroid's mass, $\vec{a}_{grav}$ is the acceleration due to gravity\footnote{Care must be taken in calculating the direction of the Earth's gravitation vector. It should be perpendicular to Earth's ellipsoid rather than towards Earth's center-of-mass; a subtle, but accumulative difference.}, $c_d$ is the drag coefficient, $A$ is the shape-density parameter \citep{bronshten_physics_1983}, $\rho_m$ is the meteoroid's density, $\rho_a$ is the atmospheric air density\footnote{The atmospheric density, $\rho_a$, is calculated using the NRLMSISE-00 empirical atmospheric model \citep{picone_nrlmsise-00_2002}.}, $c_h$ is the heat-transfer coefficient, and $H^*$ is the enthalpy of sublimation.

However, not every unknown parameter from Eq.~\ref{eq:dv_orig} and Eq.~\ref{eq:dm_orig} can be resolved as many terms are dynamically coupled, and therefore indistinguishable given only the line-of-sight measurements we obtain. Therefore, we can alter these equations by grouping the coupled terms together as shown: 

\begin{equation}\label{eq:dv/dt}
    \frac{d\vec{v}}{dt} = - \frac{\rho_a \big\|\vec{v}_{rel}\big\|}{2 \beta} \vec{v}_{rel} + \vec{a}_{grav}
\end{equation}
\begin{equation}\label{eq:db/dt}
    \frac{d\beta}{dt} = - \frac{\sigma \rho_a \big\|\vec{v}_{rel}\big\|^3}{6}
\end{equation}

where ${\beta = \sqrt[3]{m \rho_m^2} / (c_d A) = m / (c_d S)}$ is the meteoroid's ballistic coefficient, ${\sigma = c_h / (c_d H^*)}$ is the meteoroid's ablation coefficient, ${\vec{v}_{rel} = \vec{v} - \vec{v}_{atm}}$ is the meteoroid's velocity relative to the atmosphere, ${\vec{v}_{atm} = \vec{\omega}_{e} \times \vec{p}}$ is the velocity of the atmosphere, $\vec{\omega}_{e}$ is Earth's rotational angular velocity, $\vec{p}$ is the meteoroid's position in the ECI frame, and $S$ is the meteoroid's cross-sectional area. 

In addition to estimating the dynamic parameters ($\vec{p}$ and $\vec{v}$), by fitting the above differential equations to the measurements, the DTF method can also estimate some physical parameters, including the ballistic parameter, $\beta$, and ablation coefficient, $\sigma$. By assuming a constant value of the meteoroid's shape and density throughout its luminous trajectory, the fitted ballistic parameter, $\beta$, can be used to estimate the meteoroid's mass during its observed decent through the atmosphere - see Section~\ref{ssec:procedure} for details.

Although some fireball networks have sub-millisecond timing precision on their shutter actuations within a long-exposure image, such as those observatories within the GFO \citep{howie_submillisecond_2017}, the identification of the exact shutter breaks is not as precise due to halo-ing and/or saturation of the fireball. To determine any missing temporal information due to these effects, provided at least one sensor has timing for reference, the DTF method is able to handle observations without timing all-together. Additionally, the DTF method can resolve for any timing offsets between sensors, which is necessary for those meteor and fireball networks that only record relative time information.

One extra feature available as a consequence of the DTF approach is the option to include fragmentation events, which can be user-diagnosed by large flares in the light-curve. If prompted, the DTF method can resolve for both time and amount of discrete fragmentation using the deceleration characteristics of the meteoroid inherent in the observations.



\subsection{Procedure}\label{ssec:procedure}
A conceptual overview of the Dynamic Trajectory Fit (DTF) methodology will be presented here, with sufficient detail to ensure reproducibility. For reference and/or use, the Python source code will be made publicly available on the DFN's GitHub\footnote{Please follow \url{https://github.com/desertfireballnetwork/} for the source code to the DTF algorithm.}.

Computationally, the DTF algorithm is divided into three main parts: state approximation, sanity checks, and optimisation. 

\paragraph{Part 1: State Approximation}
In preparation for the main optimisation step (Part 3), we must estimate all the unknown parameters for a single point in time that describe the meteoroid's dynamics, see Eq.~\ref{eq:dv/dt} and Eq.~\ref{eq:db/dt}. The collection of these parameters is termed the meteoroid's ``state", and is given by the following vector:

\begin{equation}    
    \vec{\chi}_{est} = [p_x^f, p_y^f, p_z^f, v_x^f, v_y^f, v_z^f, \beta^f, \sigma, \delta_{frag,i}, t_{frag,i}, \Delta t_j, t_k^{rel}]
\end{equation}

where $\vec{p}^f = [p_x^f, p_y^f, p_z^f]$ is the final position, $\vec{v}^f = [v_x^f, v_y^f, v_z^f]$ is the final velocity, $\beta^f$ is the final ballistic coefficient, and $\sigma$ is the ablation coefficient. We use the end of the observable trajectory in the state estimate as it is far easier to constrain the ballistic coefficient, which relates to meteoroid mass, to be greater than zero for all times along the observable trajectory ($\beta(t)>0$). These first 8 parameters are always required to define the trajectory. If one or more fragmentation events are suspected, the percentage fragmentation, $\delta_{frag}$, and the time of fragmentation, $t_{frag}$, are added to the state for every possible event. If one or more observatories are found to contain timing offsets, an estimated offset time, $\Delta t$, is added to the state for every offset observatory. Finally, if one or more observatories contain lines-of-sight without relative times, an estimated relative time, $t^{rel}$, is added for every line-of-sight that is missing timing information.

To calculate these estimates, we must first get an idea of the trajectory from simpler triangulation methods. Using a boot-strapping approach, we can build-up from the MOP \citep{ceplecha_geometric_1987} to the SLLS \citep{borovicka_comparison_1990}, from which we can then estimate most of the state parameters. The time components are estimated first to ensure we calculate the correct position, velocity, and ballistic coefficient parameters.

To determine if any observatories have a timing offset problem, we start by assuming all the sensors are offset and determine what $\Delta t$ is needed to synchronise them. This involves first designating a "master" observatory to act as a temporal anchor, chosen as the observatory with the most lines-of-sight with timing. Next we adjust the estimated timing offsets for every other observatory to minimise the differences in lengths along the SLLS line when compared to the "master", interpolating if necessary. If the estimated offset is greater than a given tolerance, say 0.05\,seconds, then the timing from that observatory is used as relative timing only, and the estimated offset is added to the state to be optimised.

All the lines-of-sight without timing are then very roughly estimated by comparing their lengths along the SLLS line to a modelled length/time function. This function is constructed by fitting a trajectory of constant velocity to the SLLS lengths along the line over time. All timeless lines-of-sight have their along-track lengths converted to relative timing, $t^{rel}$, and are subsequently added to the state estimate to be optimised. After optimisation, these lines-of-sight each produce zero along-track error, as expected. 

Now using the rough timing corrections above, we are able to more accurately estimate the meteoroid's final position and velocity from the SLLS fit. Put simply, the estimated position is merely the final triangulated point along the SLLS line, and the estimated velocity is a least-squares average velocity of the last eight SLLS triangulated positions. The ablation coefficient, $\sigma$, is initially estimated as $14 \times 10^{-9}\,s^2/m^2$ in all cases \citep{sansom_novel_2015}. The ballistic coefficient, $\beta^f$, is roughly determined by equating the SLLS trajectory length with the propagated trajectory length assuming a $\beta^f$ value. This is achieved using Brent's root-finding method on the range $\log_{10}(\beta^f)\in[1,4]$, which approximately equates to a meteoroid mass range of 0.1\,g to 100\,ton sphere of chondritic density ($3500\,kg/m^3$).

If any fragmentation is suspected by the user, one or more fragmentation times are able to be input to the algorithm and serve as the $t_{frag}$ parameter in the state estimate. The fragmentation percent, $\delta_{frag}$ is always estimated initially as 30\%, and adjusted upon optimisation.

We must note that these estimates' sole purpose is to start the optimisation sufficiently close to the global minimum to allow convergence. Once the minimisation algorithm begins, the measurements are the only things directly influencing the trajectory solution; it is not building off already processed data.

\paragraph{Part 2: Sanity Checks}
As some fireball data reduction pipelines can be almost completely automated, such as that of the GFO, there is a chance sensor data is corrupt or has been incorrectly grouped. This could occur for a variety of reasons, including calibration errors, planes and/or satellites being misidentified as meteoroids, or the rare cases where multiple simultaneous fireballs are incorrectly correlated across sensors.

To avoid triangulation errors within the optimisation routine, a variety of sanity checks need to be performed to remove erroneous data before the optimisation is attempted. All the following checks use the rough triangulation of SLLS and the state approximation (as determined in Part 1) to ensure that each observatory's triangulated observations: 

\begin{enumerate}
    \item Decrease in height over time.
    \item Change in height at roughly the same rate.
    \item Produce sufficiently low SLLS residuals.
    \item Triangulate to positions above the ground.
    \item Triangulate to positions less than 200\,km altitude.
    \item Produce a final state velocity estimate less than 200\,km/s.
\end{enumerate}

These conditions are designed to be quite extreme to prevent accidentally discarding any valid data that has happened to triangulate poorly using the SLLS procedure. If any data is found inaccurate, the first sensor to fail an above condition is eliminated and the procedure begins over from the state approximation (Part 1).

\paragraph{Part 3: Optimisation}
Now that we have an initial state estimate (Part 1) using good data (Part 2), we are now in a position to begin the trajectory optimisation. This step could be performed with any robust minimisation routine that imposes bounds on the optimised state to ensure realistic results. For reliability, we have elected to use SciPy's in-built \textit{least-squares} function that has been thoroughly tried and tested \citep{virtanen_scipy_2019}. Within this function, the Trust Region Reflective (TRF) method is chosen as it is robust and permits bounds to be set on the allowable state. We define rather generous state bounds to give the optimisation routine enough room to effectively search the state-space while at the same time keeping the resulting state physically realistic, see Table~\ref{tab:bounds}.

\begin{table}
    \caption{State boundary conditions given to the least-squares algorithm to ensure realistic results, where LB and UB stand for lower-bound and upper-bound respectively. Also, the star-symbol represents the associated estimated state parameter as determined in Part 1.}\label{tab:bounds}
    \centering
    \begin{tabular}{|c|c|c|c|c|c|c|c|}
        \hline
        State & $\vec{p}^f$ & $\vec{v}^f$ & $\beta$ & $\sigma$ & $\delta_{frag}$ & $t_{frag}$ & $\Delta t$, $t^{rel}$ \\
        Parameters & $(km)$ & $(km/s)$ & $(kg/m^2)$ & $(kg/J)$ & $(\%)$ & $(s)$ & $(s)$ \\
        \hline
        LB   & * - 40 & * - 5 & $10^{-10}$ & $3 \times 10^{-9}$ & 0   & $t_{min}$ & * - 10 \\
        UB   & * + 40 & * + 5 & $10^4$     & $3 \times 10^{-6}$ & 100 & $t_{max}$ & * + 10 \\
        \hline
    \end{tabular}
\end{table}

The chosen TRF method also offers the option for user-defined Jacobian and state step-size. For accuracy and computational speed, we provide a parallelised custom Jacobian function that uses central differencing. The step-size is defined equal to the change in state used in the Jacobian's central differencing algorithm to avoid state divergence by overshooting the bounds of Jacobian linearity.

Once the least-squares algorithm is setup and initiated, the state is propagated using Eq.~\ref{eq:dv/dt} and Eq.~\ref{eq:db/dt} to all the other observation times and subsequently converted to lines-of-sight. These predicted lines-of-sight are differenced from the observed lines-of-sight to give the angular along-track and cross-track residual components. With the help of the Jacobian to show the direction of the local (and hopefully global) minimum, the state parameters are adjusted to minimise these angular residuals, weighted by their individual astrometric uncertainties. This procedure occurs iteratively until the state does not differ significantly enough from one iteration to the next; therefore signifying that a minimum is reached and the resulting state matches the observations as closely as possible.

Now that the optimised state solution is obtained, the state errors are determined (as discussed in Section~\ref{ssec:errors}) and propagated to all the other observation times alongside the state itself before being saved to file for subsequent orbit determination and possible darkflight analysis. Various plots are then constructed using this data, see Section~\ref{ssec:case_study}.

\subsection{Notes on Errors}\label{ssec:errors}
We must note that the least-squares algorithm used within the DTF method does not produce errors. Instead, covariance errors can be estimated afterwards from both the Jacobian of the optimised state and the covariance on the line-of-sight measurements as follows \citep{bevington_data_1993}:

\begin{equation}\label{eq:cov}
    \vec{\chi}_{cov} = \vec{\chi}_{cov}^{res} + \vec{\chi}_{cov}^{z}
\end{equation}
\begin{equation}
    \vec{\chi}_{cov}^{res} = (d\vec{\chi}/d\vec{res})^T diag(\vec{res}^2) (d\vec{\chi}/d\vec{res})
\end{equation}
\begin{equation}
    \vec{\chi}_{cov}^{z} = (d\vec{\chi}/d\vec{res})^T (d\vec{res}/d\vec{z})^T \vec{z}_{cov} (d\vec{res}/d\vec{z}) (d\vec{\chi}/d\vec{res})
\end{equation}
\begin{equation}
    d\vec{\chi}/d\vec{res} = (\vec{J} \vec{J}^T)^{-1} \vec{J}^T
\end{equation}

where $\vec{J}$ is the state Jacobian matrix, describing how the residuals change with a change in state; $d\vec{\chi}/d\vec{res}$ is the inverse of the Jacobian, describing how the state changes with a change in residuals; $d\vec{res}/d\vec{z}$ is a coordinate transform, describing how the residuals (along-track/cross-track) change with a change in line-of-sight measurements (ra/dec); $\vec{z}_{cov}$ is the covariance on the measurements; and $diag(\vec{res}^2)$ is the residual vector at the optimised state, diagonalised.

As shown in Eq.~\ref{eq:cov}, we are able to incorporate the residual covariance due to the spread in residuals around the model, $\vec{\chi}_{cov}^{res}$, and measurement covariance due to the astrometric uncertainty, $\vec{\chi}_{cov}^{z}$, into an overall covariance estimate. Separate testing showed that the measurement covariance component accurately reflected the covariance of the state through repeated Monte-Carlo analyses in which the measurements were varied within their astrometric covariance space.

However, we must also make note that the uncertainty formulation discussed above does not account for errors arising due to the meteoroid equations of motion as well as assumptions made within this model  (Eq.~\ref{eq:dv/dt} and Eq.~\ref{eq:db/dt}), such as a constant ablation coefficient, shape, and density of the meteoroid throughout the visible trajectory. Therefore, the determined covariance from Eq.~\ref{eq:cov} can be viewed as minimum uncertainties given the observations.

\section{Results and Discussion}
To demonstrate and compare the capabilities of the four previously discussed triangulation methods, we conduct two independent comparative analyses: The first study uses over 100,000 randomly simulated trajectories, comparing the fitted initial velocity vector to the simulated "truth". The second study uses a real fireball event, captured by multiple observatories within the Desert Fireball Network.

\subsection{Randomised Simulations}\label{ssec:simulations}
To fully analyse the accuracy of a triangulation algorithm through the full range of possible trajectory conditions, one must rely on simulation. Simulations allow us to compare a triangulation solution against the unaltered trajectory "truth". For the following comparative analysis, a fireball simulator was designed, built, and heavily tested under a variety of initial conditions before being used to compare the various triangulation methods. 

This fireball simulator begins with a set of randomised physical and dynamical initial conditions at the top of the atmosphere, that completely defines a meteoroid's state at that point. This randomised state is then numerically propagated forward in time using the meteoroid's 3D differential equations of motion until the meteoroid's speed relative to the ground falls below 2\,km/s. Likewise, the initial meteoroid's state is also propagated back in time until the meteoroid's height exceeds 200\,km. 

Once this simulated trajectory has been established, perfect azimuth and elevation measurements are generated every 0.1\,seconds for two (or more) randomised observatory locations for the section of the trajectory that would be visible to the sensor - i.e. while the meteoroid is more than $10^{\circ}$ above the horizon and ablating rapidly enough to be detectable from each observatory's perspective. These resulting measurements are then varied within some randomised Gaussian measurement error to better reflect reality\footnote{This trajectory can also be effected by multiple randomised or user defined fragmentation events, and/or systematic observatory timing offsets to increase realism, however these abilities are not used in this analysis.}. 

The initial state of these simulated trajectories was generated with a fixed latitude of $0^{\circ}$, a fixed longitude of $0^{\circ}$, a fixed height of $100\,km$, a uniformly random slope between $10^{\circ}$ and $90^{\circ}$, a uniformly random bearing between $0^{\circ}$ and $360^{\circ}$, and a uniformly random speed between $12\,km/s$ and $72\,km/s$. Additionally, the meteoroid was initialised with a fixed density of $3500\,kg/m^3$, a fixed spherical shape, and a uniformly random mass (in log-space) between $100\,g$ and $100\,kg$. Two uniformly random observatory locations were generated that could view the centre of the observable trajectory at an elevation greater than $20^{\circ}$. This did not always generate geometrically favourable observation combinations.

The simulated line-of-sight observations were given measurement error of 2.4\,arcmin, characteristic of the measurement errors given by a Desert Fireball Network observatory \citep{howie_how_2017}. \citet{gural_new_2012} found that the resulting radiant error was proportional to measurement error, and therefore any results found through this analysis can be linearly extrapolated to imaging systems of higher or lower resolution.  

In this analysis, we generated 123,337 sets of realistic double-station measurements from random trajectories using the fireball simulator. Each measurement set was subsequently passed to the four triangulation methods for trajectory fitting: the Method of Planes (MOP), the Straight Line Least Squares (SLLS) method, the Multi-Parameter Fit\footnote{\citet{gural_new_2012} states that "the algorithm is not ill-conditioned to having too many velocity velocity fitting parameters as long as there is measurement sample support." Therefore, we have chosen to use the exponentially dependent deceleration model specified in Eq.~4 of \citet{gural_new_2012} for MPF analysis within this paper.} (MPF) method, and the novel Dynamic Trajectory Fit (DTF) method. The original simulated radiant velocity vector and the four fitted radiant velocity vectors from the top of the trajectory are then compared, distinguishing the differences in slope, bearing, and velocity magnitude components.

Similar to the analysis performed by \citet{gural_new_2012}, the difference between the true and estimated radiant parameters are statistically analysed by considering its median value within small, equally divided bins that subtend the x-axis. This avoids excess clutter and highlights the general trends of the various triangulation methods. 

\begin{figure}
\centering
\begin{minipage}{.5\textwidth} 
    \centering
    \includegraphics[width=0.95\linewidth]{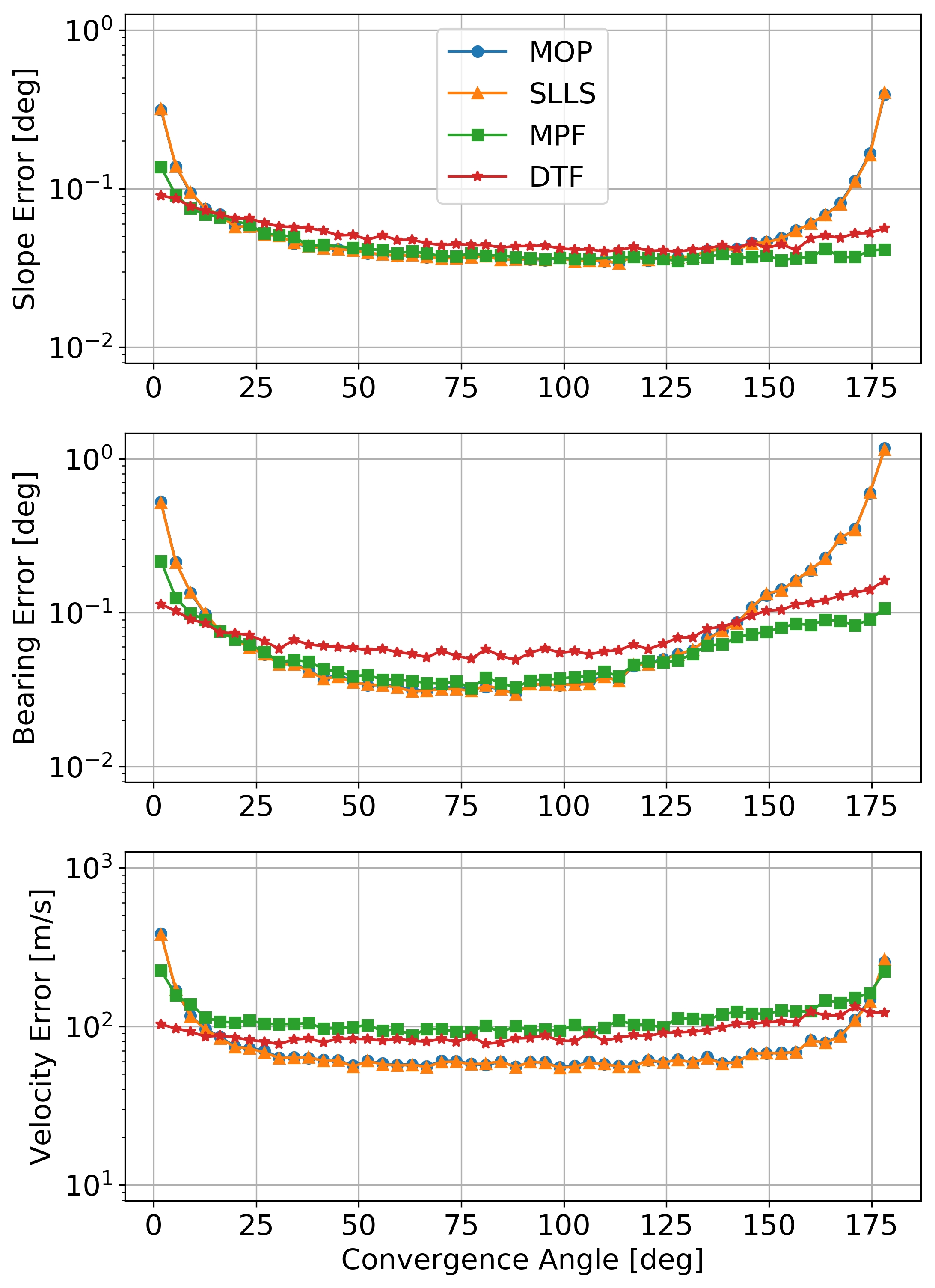}
    \captionsetup{width=0.9\linewidth}
    \captionof{figure}{The median absolute differences between the simulated and the fitted radiant at the top of the meteoroid's trajectory by varying observation convergence angle.}
    \label{fig:convergence_angle}
\end{minipage}%
\begin{minipage}{.5\textwidth} 
    \centering
    \includegraphics[width=0.95\linewidth]{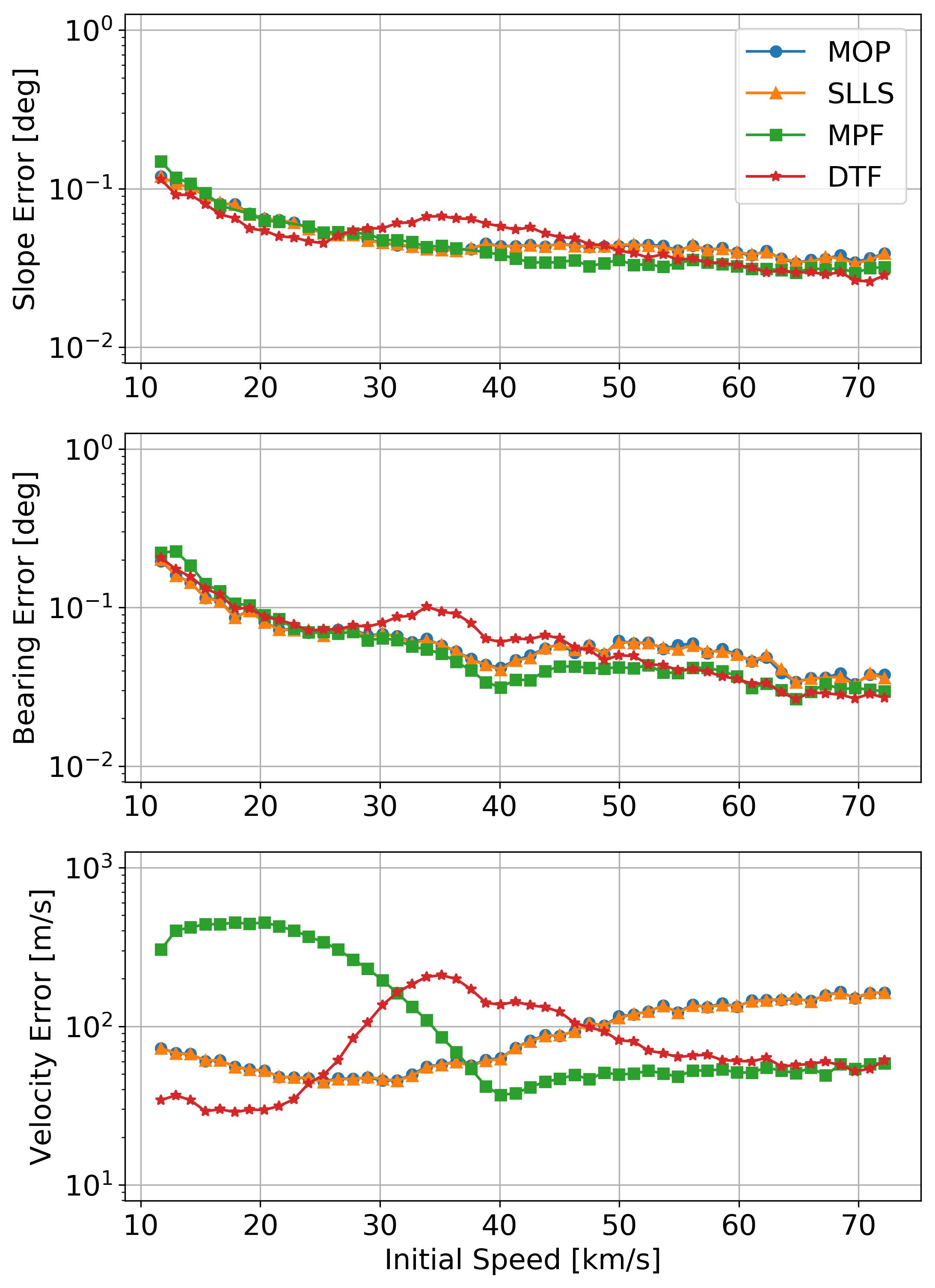}
    \captionsetup{width=0.9\linewidth}
    \captionof{figure}{The median absolute differences between the simulated and the fitted radiant at the top of the meteoroid's trajectory by varying initial speed.}
    \label{fig:initial_speed}
\end{minipage}
\end{figure}

\begin{figure}
\centering
\begin{minipage}{.5\textwidth} 
    \centering
    \includegraphics[width=0.95\linewidth]{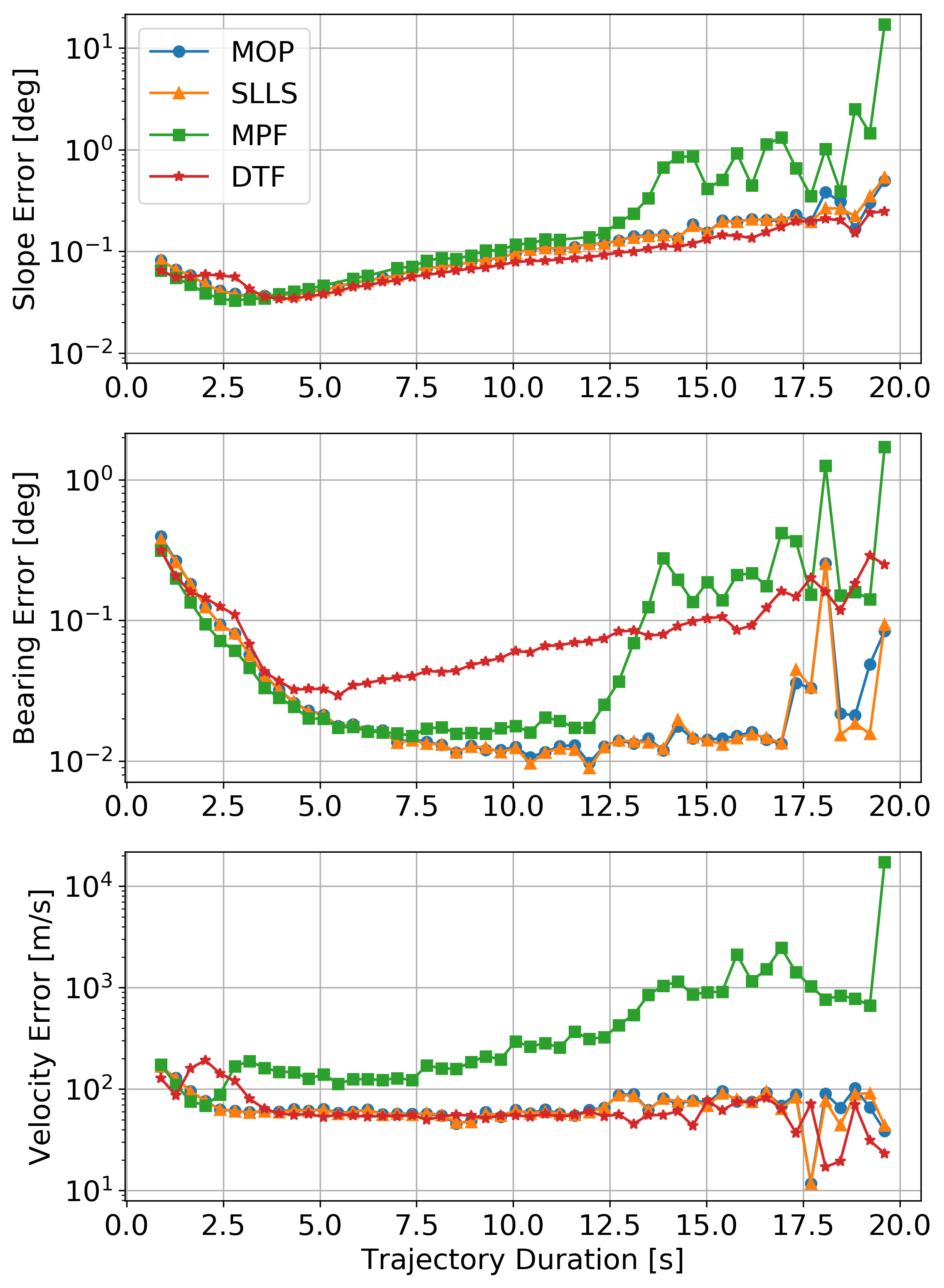}
    \captionsetup{width=0.9\linewidth}
    \captionof{figure}{The median absolute differences between the simulated and the fitted radiant at the top of the meteoroid's trajectory by varying trajectory duration, where duration is roughly proportional to number of collected observations.}
    \label{fig:trajectory_duration}
\end{minipage}%
\begin{minipage}{.5\textwidth} 
    \centering
    \includegraphics[width=0.95\linewidth]{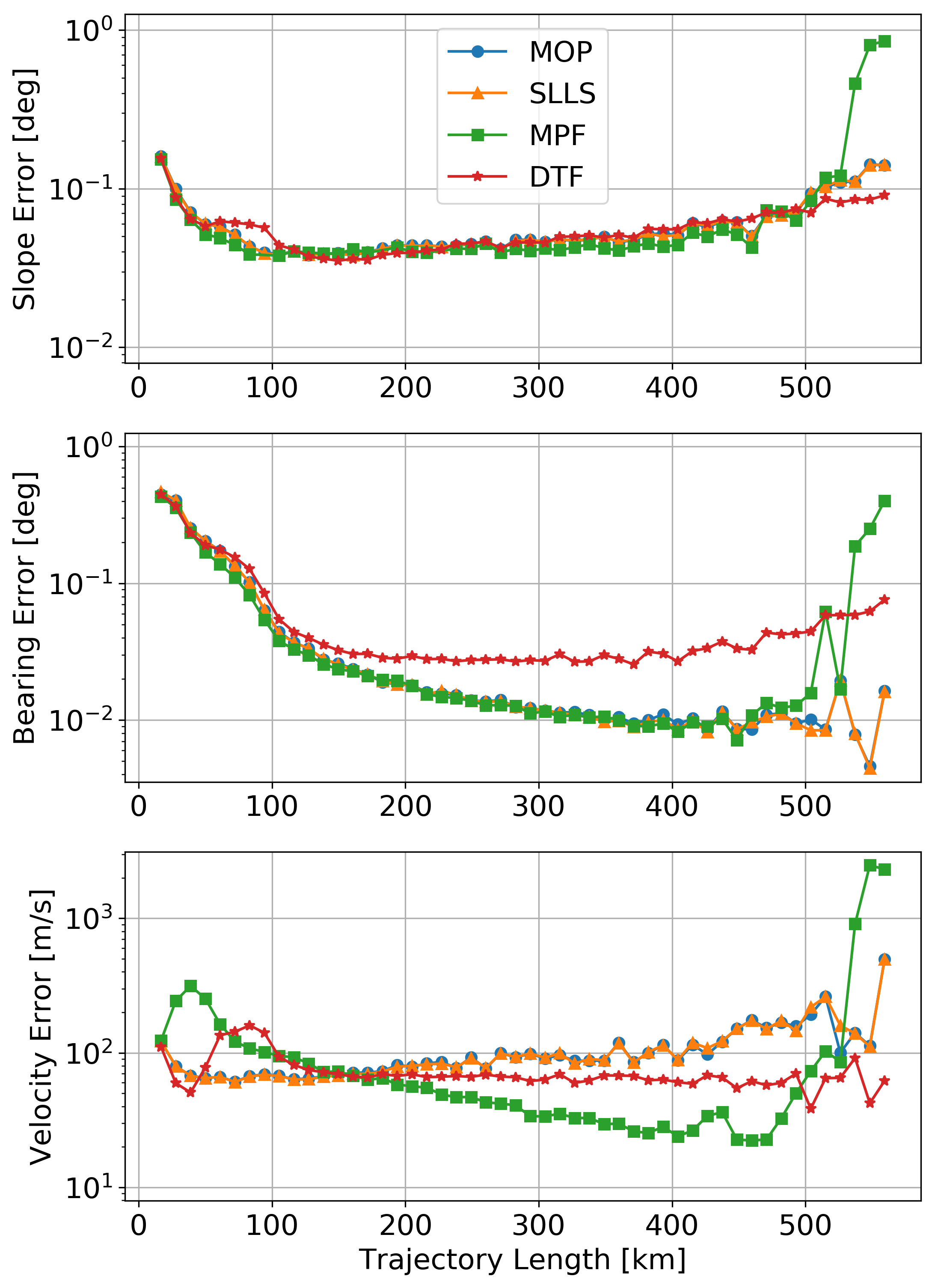}
    \captionsetup{width=0.9\linewidth}
    \captionof{figure}{The median absolute differences between the simulated and the fitted radiant at the top of the meteoroid's trajectory by varying trajectory length, where length can be roughly related to initial speed and trajectory duration.}
    \label{fig:trajectory_length}
\end{minipage}
\end{figure}

Using the approach described above, we can compare the fitting errors against different meteoroid trajectory parameters, such as observation convergence angle, initial speed, trajectory duration, and trajectory length as shown in Fig.~\ref{fig:convergence_angle}, Fig.~\ref{fig:initial_speed}, Fig.~\ref{fig:trajectory_duration}, and Fig.~\ref{fig:trajectory_length}, respectively.

From these simulation results, we notice that all triangulation methods generally agree and tend to follow the same trends. Areas of most model inaccuracy arise when a meteoroid trajectory is viewed from observatories of low convergence angle, is short in length, or displays a relatively slow entry velocity. Interestingly, these are the regions that the DTF method either matches or exceeds in accuracy when compared to the alternative triangulation methods. In particular, the DTF provides a more accurate trajectory solution at low convergence angles ($<10^{\circ}$), slow to moderate entry velocities ($<25\,km/s$), and extremely fast entry velocities ($>65\,km/s$).

Regions where the DTF method appears to perform poorly could be due to the underlying least-squares algorithm either reaching a non-global minimum or simply terminating optimisation procedures too early. Regardless, the estimated errors calculated as part of the DTF procedure (Section~\ref{ssec:errors}) are on the same order as the median absolute deviations shown from these simulations. This indicates that the true meteoroid trajectory is accurately encompassed within the DTF errors, which is the ultimate goal of meteoroid trajectory modelling.

It is also interesting to note that in most trajectory scenarios, the modelled velocity error is on the order of 0.1\,km/s. However, as stated before in \citet{gural_new_2012}, the magnitude of this model error is directly proportional to the uncertainty in the line-of-sight observations. Therefore, we can conclude that meteoroid events with observation errors less than the 2.4\,arcmin simulated here should result in a velocity accuracy better than $\sim$0.1\,km/s - the threshold needed for accurate identification of meteoroid source regions within the Solar System \citep{granvik_identification_2018}.

\subsection{Case Study: Fragmentation Event (DN141125\_01)}\label{ssec:case_study}
\begin{figure} 
    \centering
    \includegraphics[width=\linewidth]{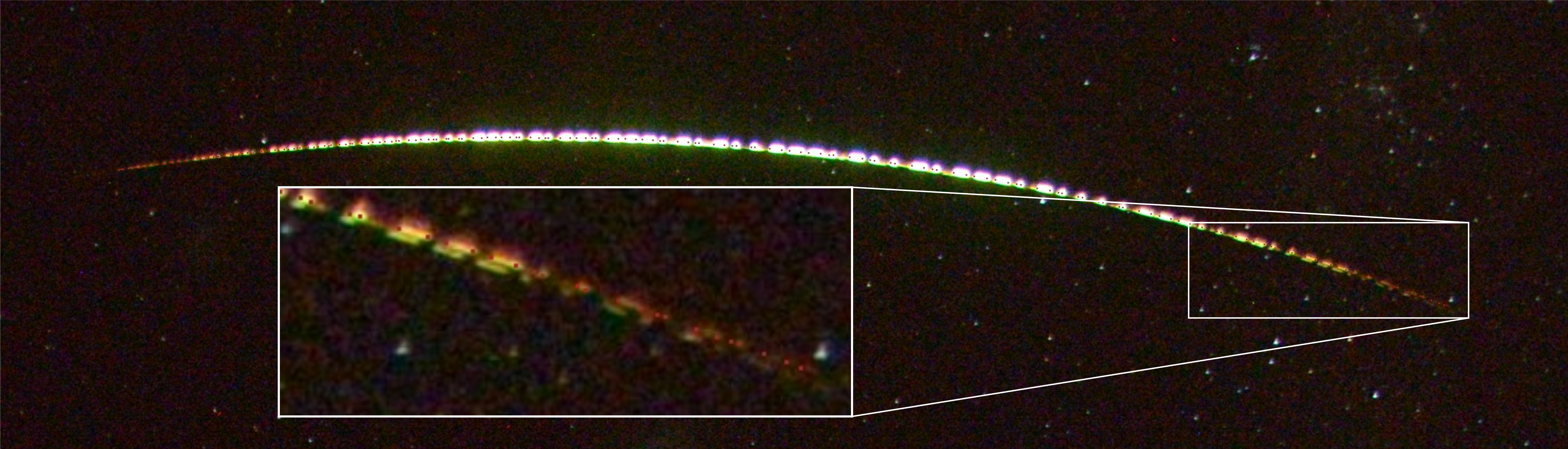}
    \caption{The captured long-exposure image of event DN141125\_01 taken from the Mulgathing station within the Desert Fireball Network, showing visible signs of fragmentation towards the end of the luminous trajectory.}
    \label{fig:fragmentation_image}
\end{figure}

Simulations are a way to thoroughly investigate and compare various models to an estimated reality. However, no simulation can 100\% replicate reality. It is for this reason that we analyse and compare the various meteoroid triangulation methods using a real-world example. We choose an event with visible signs of fragmentation to highlight the fragmentation handling within the DTF method, as shown in Fig.~\ref{fig:fragmentation_image}.

\begin{figure} 
    \centering
    \includegraphics[width=\linewidth]{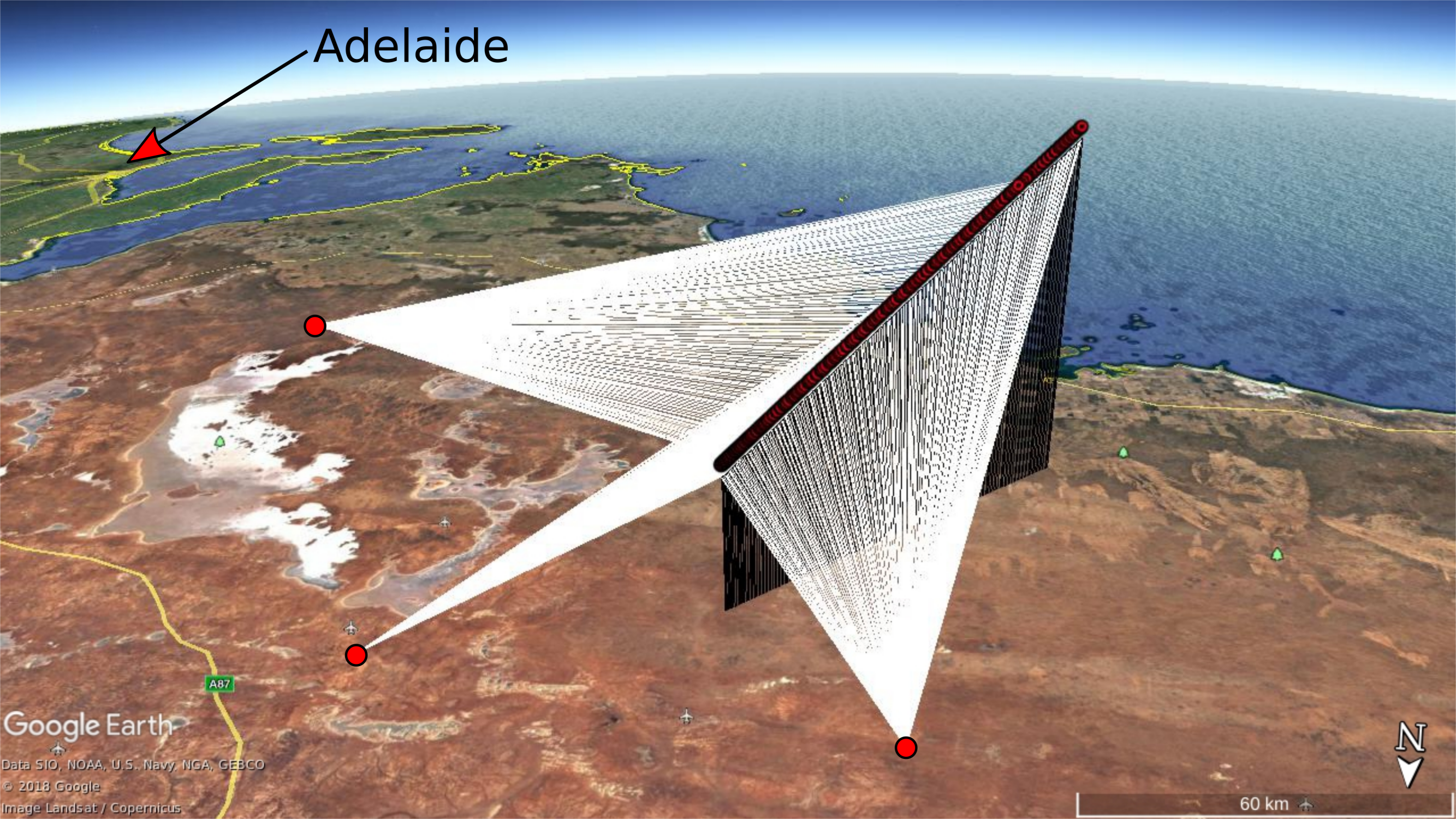}
    \caption{The triangulation of event DN141125\_01 using a total of 459 line-of-sight measurements from three South Australian observatories within the Desert Fireball Network: Mulgathing, Northwell, and Mount Ives.}
    \label{fig:fragmentation_triangulation}
\end{figure}

This fireball event with visible fragmentation, referred to as DN141125\_01, was captured by five DFN observatories - two of which could not be resolved for timing due to the distance of the observations. Although the DTF method can incorporate data with this lack of timing information, we chose to discard the data from these observatories for triangulation comparison purposes. The DN14125\_01 event was visible for 9.24\,seconds, comprising of 459 line-of-sight observations at a maximum convergence angle of $35^{\circ}$. The triangulation for event DN141125\_01 is shown visually in Fig.~\ref{fig:fragmentation_triangulation} and is summarised in Table~\ref{tab:tri_comparisionss}. 

\begin{table}
    \caption{Summary of event DN141125\_01 trajectory parameters using the four triangulation methods discussed in this paper: the Method of Planes (MOP), the Straight Line Least Squares (SLLS), the Multi-Parameter Fit (MPF), and the Dynamic Trajectory Fit (DTF). In addition, the triangulation solution of the Dynamic Trajectory Fit with fragmentation (DTF$_{frag}$) was also given to highlight this added fitting feature. The results are divided into four sections; the standard deviations of the trajectory residuals to indicate goodness of fit, the radiant direction for possible meteor stream classification, the initial trajectory position and velocity at 15:21:15.386\,UTC used for orbit determination, and the final trajectory position and velocity at 15:21:24.626\,UTC used for darkflight analysis.} 
    \centering
    \begin{tabular}{|l|c|c|c|c|c|c|c|c|}
        \hline
         & \textbf{MOP} & \textbf{SLLS} & \textbf{MPF} & \textbf{DTF} & \textbf{DTF$_{frag}$} \\
        \hline
        ATR [']    & 8.700 & 4.099 & 6.385 & 2.543 & 2.332 \\
        CTR [']    & 2.427 & 0.861 & 2.314 & 3.392 & 3.411 \\
        Total [']  & 9.033 & 4.188 & 6.792 & 4.240 & \textbf{4.132} \\
        \hline
        RA$_{\infty}$ [$^{\circ}$]  & 345.257 & 345.101 & 345.088 & 345.014 & 345.010 \\
        Dec$_{\infty}$ [$^{\circ}$] & -46.398 & -46.701 & -46.663 & -46.333 & -46.311 \\
        \hline
        Lat$_0$ [$^{\circ}$]        & -31.593 & -31.600 & -31.600 & -31.593 & -31.592 \\
        Lon$_0$ [$^{\circ}$]        & 133.770 & 133.767 & 133.765 & 133.768 & 133.769 \\
        Hei$_0$ [$km$]              &  80.441 &  80.752 &  80.815 &  80.285 &  80.189 \\
        Vel$_0$ [$\frac{km}{s}$]    &  13.977 &  14.095 &  14.381 &  13.989 &  13.908 \\
        Slope$_0$ [$^{\circ}$]      &  26.710 &  26.712 &  27.236 &  26.532 &  26.520 \\
        Azi$_0$ [$^{\circ}$]        & 230.046 & 229.688 & 228.656 & 230.054 & 230.075 \\
        Mass$_0$ [$kg$]             &    N/A &     N/A &     N/A &    0.901 &   1.605 \\
        \hline
        Lat$_f$ [$^{\circ}$]        & -31.011 & -31.011 & -31.012 & -31.010 & -31.010 \\
        Lon$_f$ [$^{\circ}$]        & 134.545 & 134.541 & 134.538 & 134.539 & 134.540 \\
        Hei$_f$ [$km$]              &  30.456 &  30.627 &  30.732 &  30.543 &  30.521 \\
        Vel$_f$ [$\frac{km}{s}$]    &   4.711 &   4.954 &   3.041 &   4.738 &   4.892 \\
        Slope$_f$ [$^{\circ}$]      &  26.705 &  26.707 &  27.232 &  25.822 &  25.861 \\
        Azi$_f$ [$^{\circ}$]        & 230.043 & 229.686 & 228.654 & 231.889 & 231.803 \\
        Mass$_f$ [$kg$]             &     N/A &     N/A &     N/A &   0.081 &   0.113 \\
        \hline
    \end{tabular}
    \label{tab:tri_comparisionss}
\end{table}

\begin{figure} 
    \centering
    \includegraphics[width=\linewidth]{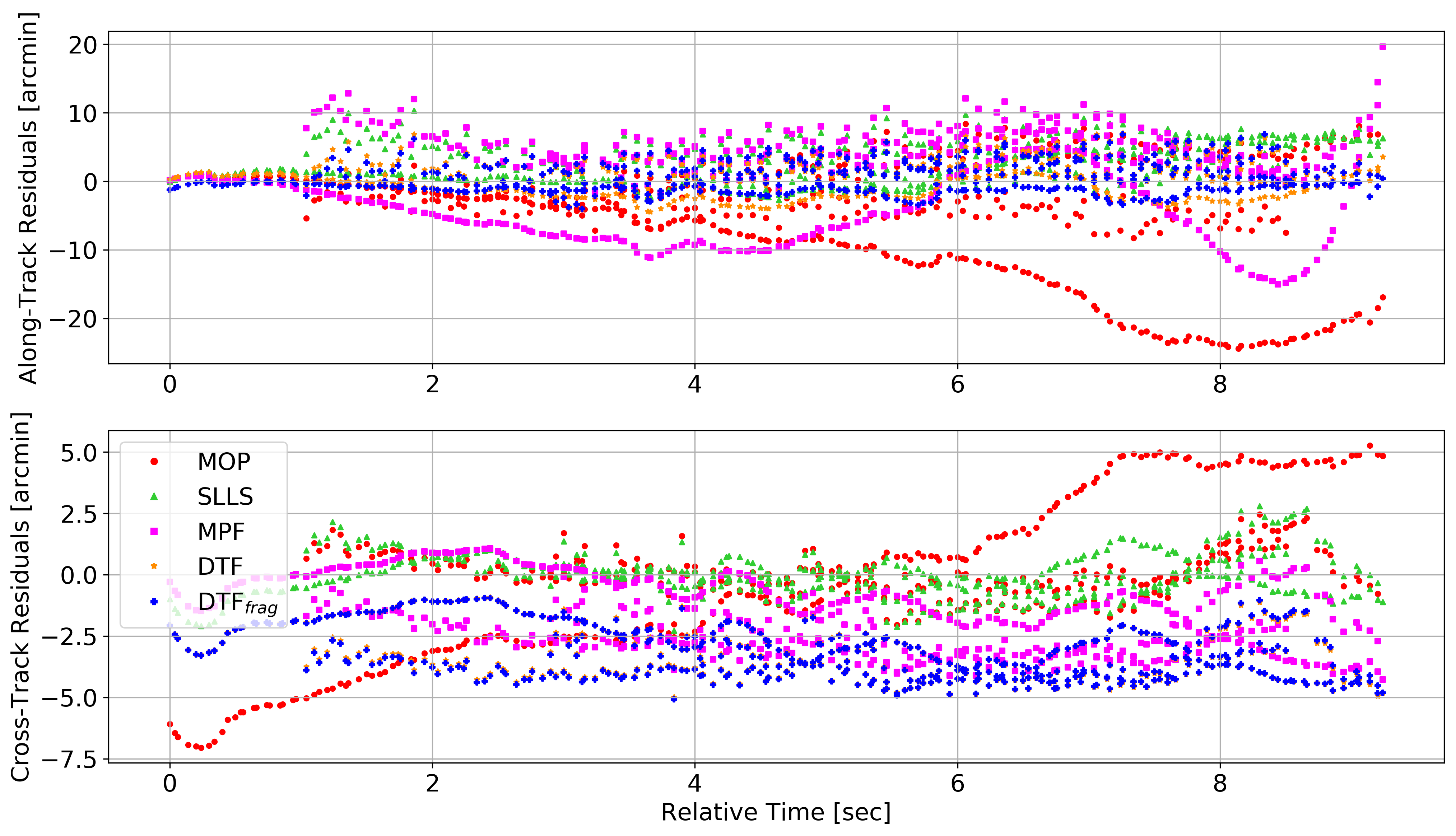}
    \caption{The along-track and cross-track residuals from all three observatories of the DN141125\_01 event using five triangulation methods: the four discussed methods as well as the Dynamic Trajectory Fit method with fitted fragmentation (DTF$_{frag}$). The standard deviation of these residuals are given in Table~\ref{tab:tri_comparisionss}.}
    \label{fig:residual_comparison}
\end{figure}

To determine which triangulation model best fits the line-of-sight observations, we compare the residual magnitudes as stated in Table~\ref{tab:tri_comparisionss} and shown more thoroughly in Fig.~\ref{fig:residual_comparison}. Unsurprisingly, the residuals in the cross-track direction are smallest using the SLLS method as this is its optimisation parameter. However, the DTF$_{frag}$ model possesses the smallest total residuals.

\begin{figure} 
    \centering
    \includegraphics[width=\linewidth]{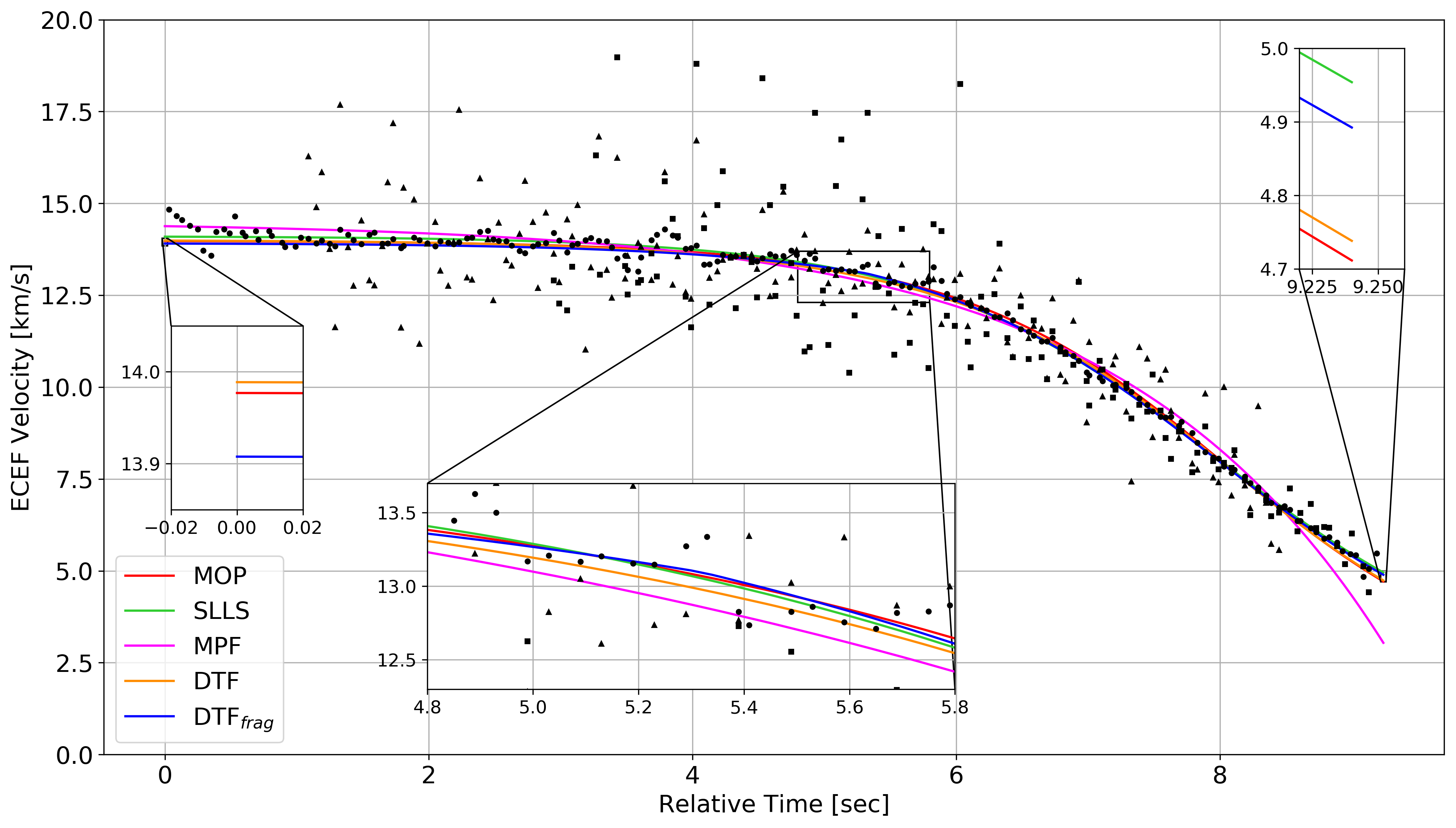}
    \caption{The modelled velocity of the DN142511\_01 event from the various triangulation methods, as discussed in Section~\ref{sec:intro2} and Section~\ref{sec:dtf}. The surrounding scatter is the instantaneous velocities as calculated by the change in adjacent SLLS positions along the trajectory over the change in time from each observatory, separately. Velocity subtleties at the beginning, middle, and end of the trajectory are highlighted by zoomed in sections.}
    \label{fig:velocity_comparison}
\end{figure}

The velocities determined by the various triangulation methods rely on different models, each containing unique assumptions. The velocity determination algorithm used within the MOP and SLLS methods fits the 1D meteoroid equations of motion to the lengths along the 1D trajectory, assuming an exponential atmosphere \citet{pecina_new_1983}. The velocity calculated by the MPF method uses a purely empirical formula \citep{whipple_reduction_1957, gural_new_2012}. Lastly, the velocity results from the DTF method consults the meteoroid's 3D equations of motion directly, without any simplifying straight line or atmospheric assumptions. The subtleties between these velocity models using data from event DN141125\_01 are compared in Fig.~\ref{fig:velocity_comparison}.

As shown in Table~\ref{tab:tri_comparisionss} and Fig.~\ref{fig:velocity_comparison}, the final velocity predicted by the MPF method does not appear to follow the instantaneous velocity scatter, suggesting the exponentially dependent velocity model does not reflect reality for long fireball-type events. Excluding the MPF velocity, the remaining velocity models seem very similar, varying by about 300\,m/s at the extremities. However, this 300\,m/s variation would still lead to considerably different darkflight and orbit regression results.

\begin{figure} 
    \centering
    \includegraphics[width=\linewidth]{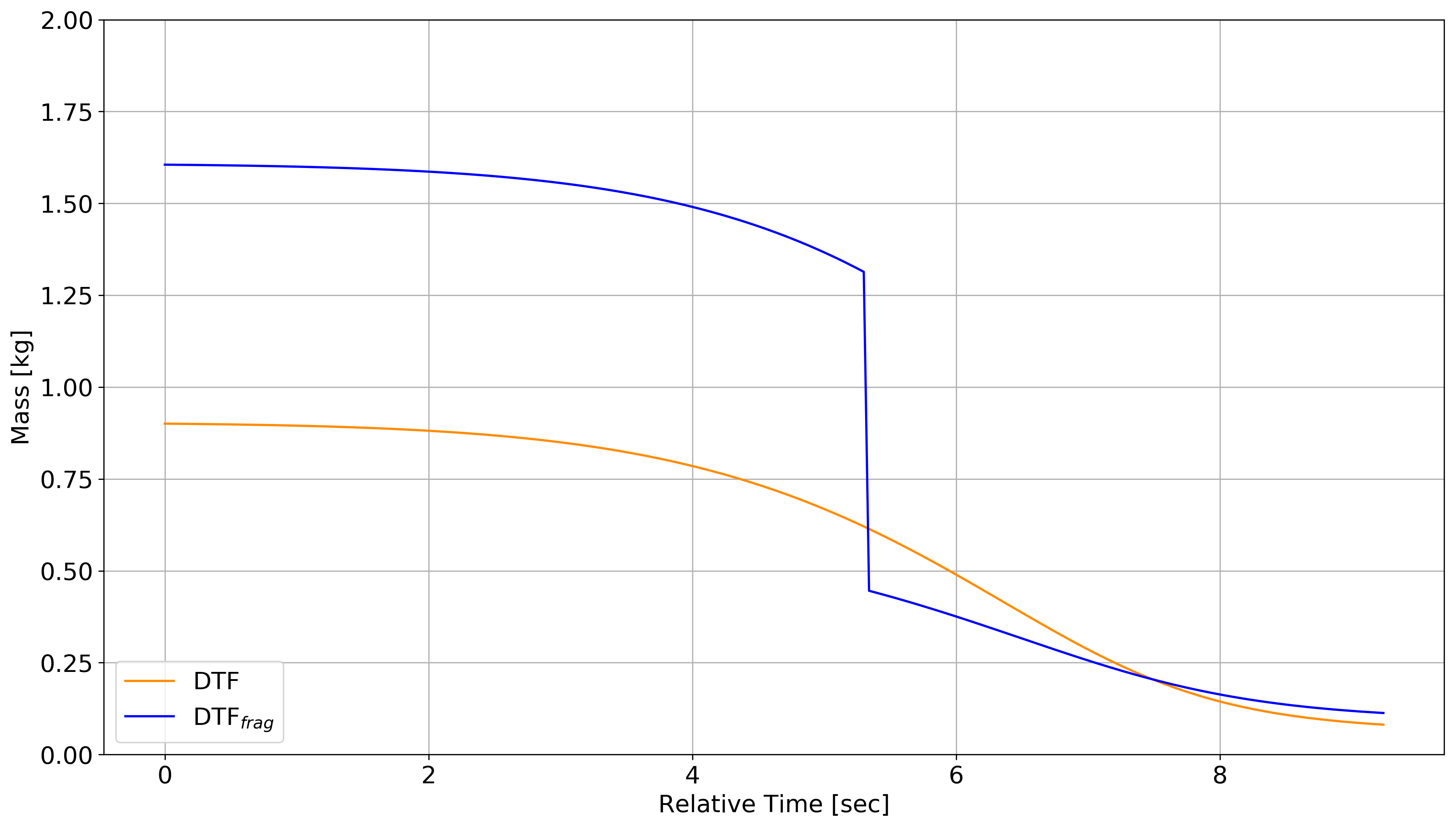}
    \caption{The estimated mass of the DN141125\_01 event throughout its trajectory determined by the Dynamic Trajectory Fit, both with (DTF) and without (DTF$_{frag}$) fragmentation fitting. The other three triangulation models are not plotted here as they do not produce mass estimates.}
    \label{fig:mass_comparisonn}
\end{figure}

As discussed in Section~\ref{sec:dtf}, the DTF method is able to resolve for the meteoroid's ballistic coefficient over time, $\beta(t)$. By assuming a constant meteoroid shape and density, we can estimate the meteoroid's mass throughout the observed luminous trajectory directly using the line-of-sight observations - unlike any other compared triangulation method. This feature not only helps diagnose meteorite-dropping events, but assists greatly in constraining the meteorite search area. The mass estimates for event DN141125\_01 using the DTF and DTF$_{frag}$ methods are compared in Fig.~\ref{fig:mass_comparisonn}. The DTF$_{frag}$ method predicts the meteoroid from DN141125\_01 broke up around 5.3\,seconds, at an altitude of 47.3\,km - consistent with the visible fragmentation shown in Fig.~\ref{fig:fragmentation_image}.

To summarise this comparison of triangulation methods, the Dynamic Trajectory Fit with fragmentation handling (DTF$_{frag}$) appears to be the best model for event DN141125\_01. While there may be events that are better suited to the other triangulation methods, the simulations discussed in Section~\ref{ssec:simulations} show that the DTF method is an equal if not better choice for most events. Additionally, the DTF method can estimate the mass of the meteoroid from the line-of-sight observations directly, as discussed in Section~\ref{ssec:mop}, Section~\ref{ssec:slls}, and Section~\ref{ssec:mpf}. 



\section{Future Functionality}
While the proposed DTF method appears successful in its current form, and poses considerable merit, there are a few improvements that could be applied to increase realism and draw out additional subtleties within the gathered data. These improvements include:

\paragraph{Light-curve incorporation} With the inclusion of light-curve data, we would have the opportunity to better model meteoroid mass-loss along the trajectory, which would act to further constrain the meteoroid state and its associated uncertainty. Luminous efficiency models, such as \citet{gritsevich_constraining_2011}, could be relatively easily incorporated into the state propagation of the meteoroid to better estimate its physical and dynamical parameters.

\paragraph{Automated fragmentation determination} Currently, we rely on a user-defined time of fragmentation. However, with full light-curve history, we should be able to flag fragmentation events from light-curve peaks alone, therefore negating any user-required input to the algorithm. However, this functionality could easily be integrated upstream in a larger data reduction pipeline using measurements from highly sensitive radiometers \citep{buchan_developing_2019}, not necessarily integrated in the triangulation method itself.


\paragraph{Meteoroid spin modelling} For some particularly long fireballs, such as \textit{Case~1} of \citet{sansom_3d_2019}, trajectories appear to considerably deviate from the fall-plane, suggesting there are unaccounted aerodynamic effects. We hypothesise this might be in part due to the Magnus Effect at high velocities; that is, the resulting curvature of an objects trajectory due to its spin. It would be very interesting to model these cases with meteoroid spin considered. The proposed DTF method would simply require an additional three state parameters to model this phenomena; namely the angular velocity vector, $\vec{\omega}_{spin} = [\omega_x, \omega_y, \omega_z]$. It would be conceivable to extend the DTF method to optimise without spin, only re-optimising with spin if the measurements did not adequately match the model (reduced chi squared $\chi_{\nu}^2\approx1$).  

\section{Conclusions}
Meteoroid orbits and meteorite samples provide invaluable information that helps planetary scientists investigate Solar System formation and the origin of life on Earth. Fireball networks around the globe are on the forefront of providing this knowledge. However, the accuracy of the determined orbit and the chance of meteorite recovery both rely heavily on the accuracy of the underlying meteoroid triangulation method.

Three triangulation methods have been proposed in the past: the Method of Planes \citep{ceplecha_geometric_1987}, the Straight Line Least Squares method \citep{borovicka_comparison_1990}, and the Multi-Parameter Fit \citep{gural_new_2012}. The first two listed methods above separate out the geometric fit from the dynamic modelling. In 2012, Gural simplified this procedure to a single step, changing the well-known convergence angle from that between planes to that between simultaneous rays - a clear advantage over the past traditional triangulation methods. However, the velocity models suggested within \citet{gural_new_2012} are empirically derived for small meteors and do not reflect reality, particularly for meteorite-dropping events. The proposed novel Dynamic Trajectory Fit method not only contains a more realistic dynamic model, but it possesses the ability to determine the meteoroid's ballistic coefficient throughout the observable trajectory directly from the line-of-sight measurements - unlike any other proposed triangulation method. With meteoroid shape and density assumptions, this ballistic coefficient can be easily translated into meteoroid mass.

Over 100,000 multi-station meteoroid simulations revealed the advantage of the Dynamic Trajectory Fit method particularly for relatively slow entry events ($<$25\,km/s) as well as events observed from low convergence angles ($<$10$^{\circ}$). Additionally, a visibly fragmenting fireball event captured by three stations of the Desert Fireball Network was used to compare the four triangulation methods. The Dynamic Trajectory Fit with fragmentation was shown to best match the observations, with the predicted fragmentation time in agreement with the observed data.

The method proposed here could be easily modified to fit arbitrarily complex equations of motion, to include light-curve data, and to provide automated fragmentation detection in the future.

\section{Acknowledgements}
This work was funded by the Australian Research Council as part of the Australian Discovery Project scheme, and supported by resources provided by the Pawsey Supercomputing Centre with funding from the Australian Government and the Government of Western Australia. This work was also supported by an Australian Government Research Training Program (RTP) Scholarship.

This research made use of Astropy, a community-developed core Python package for Astronomy \citep{collaboration_astropy:_2013}. Additionally, the majority of figures were generated using Matplotlib, another community-developed Python package \citep{hunter_matplotlib:_2007}.

\section{References}
    \bibliographystyle{abbrvnat}
    \bibliography{bib}

\end{document}